\documentclass[runningheads]{llncs}
\usepackage[T1]{fontenc}
\usepackage{graphicx}
\usepackage{xcolor}
\usepackage{amsmath,amssymb,amsfonts}

\usepackage{hyperref}
\usepackage{color}
\usepackage{forest}
\usepackage{tikz}
\usetikzlibrary{arrows.meta, positioning, shapes.geometric}

\definecolor{processblue}{cmyk}{0.96,0,0,0}

\usepackage[linesnumbered, noend]{algorithm2e}
\RestyleAlgo{ruled}
\SetKwProg{Function}{function}{}{}

\newcommand{\stree}{\ensuremath{\textrm{STree}}}
\newcommand{\fstree}{\ensuremath{\textrm{\tiny STree}}}

\newcommand{\cdawg}{\ensuremath{\textrm{CDAWG}}}
\newcommand{\scdawg}{\ensuremath{\textrm{SCDAWG}}}
\newcommand{\fcdawg}{\ensuremath{\textrm{\tiny CDAWG}}}
\newcommand{\fscdawg}{\ensuremath{\textrm{\tiny SCDAWG}}}
\newcommand{\len}[1]{\vert #1 \vert}

\newcommand{\rev}[1]{#1^{\text{rev}}}

\begin{document}
\title{Optimal-Time Contextual Pattern Matching \\ in Compressed Space\thanks{Funded by Basal Funds FB0001 and AFB240001, and Fondecyt Grant 1260080, ANID, Chile.}}
%
%
\author{Gonzalo Navarro\inst{1,2}\orcidID{0000-0002-2286-741X} and Francisco Olivares\inst{1,2}\orcidID{0000-0001-7881-9794}}
\authorrunning{Gonzalo Navarro and Francisco Olivares}
%
\institute{Centre for Biotechnology and Bioengineering, Chile \and 
Department of Computer Science, University of Chile, Chile\\
\email{\{gnavarro,folivares\}@uchile.cl}
}
\maketitle              
\begin{abstract}
Contextual pattern matching is the task of, given a pattern $P[1,m]$, a context length $\lambda$, and a text $T[1,n]$, find all the $occ$ distinct contexts in which $P$ occurs in $T$, the context being the $\lambda$ symbols preceding and the $\lambda$ symbols following the occurrence; a text position where each context occurs must be output. While the problem can be solved in optimal time $O(m+occ)$ using $O(n)$-space precomputed data structures on $T$, this type of search is particularly relevant on large repetitive text collections, where $O(n)$ space can be prohibitive. We present the first optimal-time solution that runs in compressed space, namely that of a symmetric CDAWG (SCDAWG) of $T$. Further, we show how the set of $occ$ solutions can be enumerated with $O(\log\log\lambda)$ delay after $O(m)$-time preprocessing of $P$. To achieve this, we develop an improved linear-space distance-sensitive weighted ancestor data structure.
\keywords{Compressed text indexing \and CDAWGs \and Contextual Pattern Matching \and Weighted ancestors in trees}
\end{abstract}

\section{Introduction}

Given a text $T[1, n]$ and a pattern $P[1, m]$, the classical Pattern Matching problem consists of finding all of the positions of $T$ where $P$ occurs. The Contextual Pattern Matching problem is a recently proposed variant \cite{Navarro2020:contextual} better suited for searching large collections of similar documents (e.g., genome collections, versioned documents, etc.). In such a case, if $P$ occurs inside a piece of text that is repeated many times along the collection, it may be wasteful to list all those occurrences individually, as these are, for many purposes, essentially the same occurrence. We might want to list, for example, all the different contexts in which some DNA motif occurs in a large genomic collection: while most occurrences can appear within identical contexts, different contexts may indicate genetic variants or diseases. As another example, we might want to track the changes on what is said about a proper name is mentioned along the different versions of a Wikipedia page by listing the different snippets around its occurrences, which are likely to be identical across many versions.  

In Contextual Pattern Matching, the query provides also a {\em context} length $\lambda$ and asks for one occurrence position of each distinct context around $P$: the context is formed by the $\lambda$ characters preceding and the $\lambda$ characters following the occurrence. This is formally stated as follows.

\begin{definition}[\cite{Navarro2020:contextual}]
The Contextual Pattern Matching problem on a text $T[1,n]$ is,
given a pair $(P[1,m],\lambda)$, return a position in $T$ for each of the $occ$ distinct strings
$XPY$ occurring in $T$, for all $X$, $Y$ such that $|X| = |Y | = \lambda$. To capture the occurrences
near the extremes of $T$, assume $T$ is preceded and followed by $\lambda$ copies of the
special symbol $\$$, which does not appear in $P$.
\end{definition}

In that paper \cite{Navarro2020:contextual}, it was shown that the problem can be solved in optimal time, $O(m+occ)$, by using $O(n)$-size data structures (essentially, suffix trees). This is not so satisfactory, however, because many repetitive collections are too large for $O(n)$ space to be affordable. Instead, one desires some {\em compressed indexing} solution, which provides search functionality within space bounded by some measure of compressibility \cite{Navacmcs20.3,Navacmcs20.2}. 

In the same paper (later corrected and improved \cite{Nav20ContextualArxiv}) they provide one such compressed indexing solution: Letting $r$ be the number of runs in the Burrows-Wheeler Transform of $T$, and $\overline{r}$ the sum of the values of $r$ for $T$ and its reverse, the problem can be solved in time $O(m + occ \log n)$ with an index of size $O(\overline{r}\log\frac{n}{\overline{r}})$. Later, Abedin et al.~\cite{AbedinCGT2023:contextual} obtained $O(m + occ\log \lambda \log\frac{n}{r})$ time and $O(r \log \frac{n}{r})$ space.
Using bidirectional indexes, the problem can be solved in time $O((m+\lambda \cdot occ)\log\log n)$, within $O(\overline{r})$ space \cite{Nav20ContextualArxiv}.

A natural question is whether one can obtain optimal time, $O(m+occ)$, within space proportional to some measure of repetitiveness (like $r$ and $\overline{r}$, in the current solutions). In this paper we present the first index of this kind, whose space is related to measure $e$, which is the size of the CDAWG of $T$ \cite{BBHMCE87}. The CDAWG is a directed acyclic graph obtained by merging identical subtrees of the suffix tree: since the suffix tree of repetitive text collections tends to have many identical subtrees, $e$ is a meaningful measure of repetitiveness. We actually build on a Symmetric CDAWG (SCDAWG) \cite{scdawg}, which includes the $\overline{e}$ edges of the $\cdawg$s of $T$ and of the reverse of $T$ (the nodes are the same for both CDAWGs). Although it always holds $\overline{e} \ge r$, $\overline{e}$ is incomparable with $r \log \frac{n}{r}$ \cite{Navacmcs20.3}, which is the best space of previous solutions \cite{Nav20ContextualArxiv,AbedinCGT2023:contextual}. Concretely, our index uses space $O(\overline{e})$ and performs Contextual Pattern Matching in optimal time, $O(m+occ)$.

We then turn our attention to the enumeration problem: since the number of occurrences can be huge, we wish to provide an efficient enumerator. This is a data structure that, after some (ideally small) preprocessing time, can deliver any new occurrence within a bounded {\em delay}. For example, it is easy to support enumeration for Pattern Matching, using a suffix tree with threaded leaves, with $O(m)$ preprocessing and $O(1)$ delay, but this is not so easy for Contextual Pattern Matching, even with $O(n)$ space. Our optimal-time $O(m+occ)$ algorithm naturally yields an enumerator with $O(\lambda)$ delay, but we can do better: We give the first enumerator for Contextual Pattern Matching that, using $O(\overline{e})$ space, provides $O(m)$ preprocessing time and $O(\log\log\lambda)$ delay.

To obtain this delay, we develop an improved linear-space distance-sensitive data structure for weighted ancestor queries \cite{FM96,KL07,BLRSRW15}, which is of independent interest. Its query time is loglogarithmic on the difference of weights between the query and answer nodes. We build our solution on top of previous work \cite{KL07}, which incurs an $O(\log^* n)$ extra penalty factor in order to achieve linear space. Using new techniques from the realm of compact data structures, we reduce this extra factor to constant (another solution to remove the $O(\log^* n)$ term exists \cite{gawrychowski2014weighted}, but we believe ours is simpler).

\section{Preliminaries}\label{sec:preliminaries}
By $[i, j]$ we denote $\{i, i + 1, \dots, j\}$, if $i \leq j$, and $[i, j] = \emptyset$, otherwise. By $[n]$ we refer to $[1, n]$. $T[1, n]$ denotes a length-$n$ string over an alphabet $\Sigma = \{1, \dots, \sigma\}$. By $\vert T \vert$ we refer to the length of $T$. $T[i]$ denotes the $i$-th symbol of $T$, and $T[i, j] = T[i] \cdots T[j]$ if $[i, j] \neq \emptyset$, otherwise ($j < i$) $T[i, j] = \epsilon$, which is the only string satisfying $\vert \epsilon \vert = 0$. $T[1, i]$ is the prefix of $T$ ending at position $i$ and $T[j, n]$ is the suffix of $T$ starting at position $j$. We assume $T$ ends with a special symbol $\$$ that only occurs at $T[n]=\$$ and that is smaller than any other symbol $c \in \Sigma$. To enforce that the special symbol $\$$ always appears at the end of our input strings, we define the reverse $\rev{T}$ of $T$ as $\rev{T} = T[n-1] \cdots T[1]\$$.

Given $T \in \Sigma^*$, we say that $T[i, j]$ is a right-maximal substring of $T$ if $j = n$ or there exists $a, b \in \Sigma$ ($a \neq b$) such that $T[i, j]a$ and $T[i, j]b$ are both substrings of $T$; $T[i, j]a$ and $T[i, j]b$ are said to be right-extensions. 
Similarly, we say that $T[i, j]$ is a left-maximal substring of $T$ if $i = 1$ or there exists $c, d \in \Sigma$ ($c \neq d$) such that $cT[i, j]$ and $dT[i, j]$ are both substrings of $T$; $cT[i, j]$ and $dT[i, j]$ are called left-extensions. 
We say that $T[i, j]$ is maximal if it is both right-maximal and left-maximal.

\subsection{Suffix trees}

The \emph{suffix tree} \cite{Wei73,McC76,Ukk95} of a string $T$ over alphabet $\Sigma$, $\stree(T) = (V^T_{\fstree}, E^T_{\fstree})$ is the compacted trie of all suffixes of $T$, where $V^T_{\fstree}$ is the set of nodes and $E^T_{\fstree}$ is the set of edges. Every node with out-degree greater than zero is called an \emph{internal node}. The remaining nodes are called \emph{leaves}. The elements of $E^T_{\fstree}$ are tuples of the form $(u, s, v)$ where the origin $u$ is an internal node, the destination node $v \neq u$ cannot be the root, and $s \in \Sigma^+$ is the label of the edge; we call $|s|$ the edge length. We denote by $\ell(v)$ the string obtained by concatenating the edges' labels in the path from the root to a node $v$. It is well-known that each $\ell(v)$ corresponds to a different right-maximal substring of $T$, and each right-maximal substring $s$ of $T$ is the label $\ell(v)$ for some $v$, so there is a bijection between nodes in $V^T_{\fstree}$ and right-maximal substrings of $T$.

From the latter we can see that every internal node has at least two outgoing edges, and the labels of all the edges leaving a node start with a different symbol. Additionally, the edges leaving a node are sorted lexicographically by their labels, so concatenating the labels of the edges in the path from the root to the $i$-th leaf spells out the $i$-th suffix of $T$ in lexicographically sorted order. Also, the leaves store the position where the suffix it represents starts in $T$ and collecting those positions in leaves from left to right we obtain the \emph{suffix array} \cite{ManberM93}.

The {\em locus} of $P$ is the highest node (i.e., whose path connecting it with the root has the least number of edges) $v$ such that $P$ is a prefix of $\ell(v)$. Having found the locus $u$ of $P$ by descending from the root matching its symbols, the positions of the $occ$ occurrences of $P$ can be reported in $O(occ)$ time by collecting all the leaves under $u$.

We assume the labels of edges $(u, s, v)$ are not stored explicitly but just the indices $i, j$ of one occurrence of $s$ in $T$ (namely, $s = T[i, j]$). Since the number of nodes and edges is proportional to the number of right-maximal substrings in $T$, then by storing the input string $T$, $\stree(T)$ consumes $O(n)$ space.

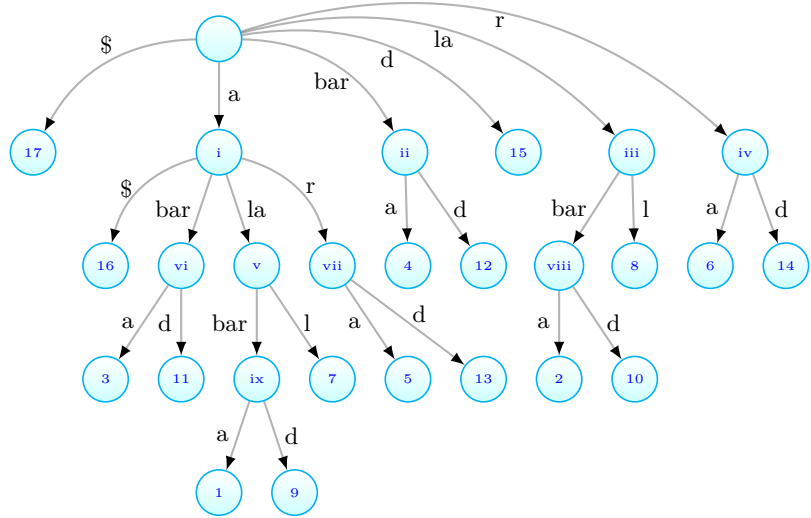
\begin{figure}[t]
\begin{center}
       \begin {tikzpicture}
         [-latex, auto, node distance = 1.5cm and 1.5cm, 
         on grid, semithick,
         state/.style = {circle, top color = white, bottom color = processblue!20,
        draw, processblue, text = blue, minimum width = .6 cm, 
        font=\tiny
        },
        every edge/.style={draw, draw opacity=0.3, thick}
        ]
      \node[state] (root) {};
      \node[state] (a) [below = of root] {i};
      \node[state] (dollar) [left = 70pt of a] {17};
      \node[state] (bar) [right = 70pt of a] {ii};
      \node[state] (d) [right = of bar] {15};
      \node[state] (la) [right = of d] {iii};
      \node[state] (r) [right = of la] {iv};

      \node[state] (ala) [below right = 1.5cm and 0.5 cm of a] {v};
      \node[state] (abar) [below left = 1.5cm and 0.5 cm of a] {vi};
      \node[state] (ar) [right = 1 cm of ala] {vii};
      \node[state] (adollar) [left = 1 cm of abar] {16};

      \node[state] (bara) [right = 1 cm of ar] {4};
      \node[state] (bard) [right = 1 cm of bara] {12};

      \node[state] (labar) [right = 1cm of bard] {viii};
      \node[state] (lal) [right = 1 cm of labar] {8};

      \node[state] (ra) [right = 1 cm of lal] {6};
      \node[state] (rd) [right = 1 cm of ra] {14};

      \node[state] (abara) [below = 1.5cm of adollar ] {3};
      \node[state] (abard) [right = 1 cm of abara] {11};

      \node[state] (alabar) [right = 1 cm of abard] {ix};
      \node[state] (alal) [right  = 1 cm of alabar] {7};

      \node[state] (ara) [right  = 1 cm of alal] {5};
      \node[state] (ard) [right  = 1 cm of ara] {13};

      \node[state] (labara) [below = 1.5cm and 0.5 cm of labar] {2};
      \node[state] (labard) [right = 1 cm of labara] {10};

      \node[state] (alabara) [below left = 1.5cm and 0.5 cm of alabar] {1};
      \node[state] (alabard) [below right = 1.5cm and 0.5 cm of alabar] {9};

      \path (root) edge[bend right] node[above] {\$} (dollar);
      \path (root) edge node {a} (a);
      \path (root) edge[bend left] node[below] {bar} (bar);
      \path (root) edge[bend left] node[below, yshift = 2pt] {d} (d);
      \path (root) edge[bend left] node[below, yshift = 3pt] {la} (la);
      \path (root) edge[bend left] node[below, yshift = 2pt] {r} (r);
      
      \path (a) edge[bend right] node[left] {\$} (adollar);
      \path (a) edge node[left] {bar} (abar);
      \path (a) edge node[right] {la} (ala);
      \path (a) edge[bend left] node[right] {r} (ar);

      \path (bar) edge node[left] {a} (bara);
      \path (bar) edge node[right] {d} (bard);

      \path (la) edge node[left] {bar} (labar);
      \path (la) edge node[right] {l} (lal);

      \path (r) edge node[left] {a} (ra);
      \path (r) edge node[right] {d} (rd);

      \path (abar) edge node[left] {a} (abara);
      \path (abar) edge node[left] {d} (abard);

      \path (ala) edge node[left] {bar} (alabar);
      \path (ala) edge node[right] {l} (alal);

      \path (ar) edge node[left] {a} (ara);
      \path (ar) edge node[above, pos = .6] {d} (ard);

      \path (labar) edge node[left] {a} (labara);
      \path (labar) edge node[right] {d} (labard);

      \path (alabar) edge node[left] {a} (alabara);
      \path (alabar) edge node[right] {d} (alabard);

    \end{tikzpicture}
  \end{center}
    \caption{Suffix tree for the string "alabaralalabarda". Edges leading to leaves show only the first symbol of their label, which continues spelling the text until the \$ terminator. Leaves are labeled with the starting position of the corresponding suffix, and internal nodes with Roman numbers.}
    \label{fig:example-suffix-tree}
\end{figure}

Figure~\ref{fig:example-suffix-tree} shows the suffix tree for the string "alabaralalabarda".

\subsection{CDAWGs}
The \emph{Compact Directed Acyclic Word Graph (CDAWG)} of $T$ \cite{BBHMCE87} is a directed acyclic graph $\cdawg(T) = (V^T_{\fcdawg}, E^T_{\fcdawg})$ that represents the suffixes of $T$. It is well-known that in $\cdawg(T)$, a node $v \in V^T_{\fcdawg}$ represents a set of nodes from $V^T_{\fstree}$, such that all of them are roots of isomorphic sub-trees of $\stree(T)$. So, if a node $u \in V^T_{\cdawg}$ represents nodes $u_1, \dots, u_k \in V^T_{\fstree}$ then $\ell(u_a)$ is a suffix of $\ell(u_b)$, for any $\len{\ell(u_a)} \leq \len{\ell(u_b)}$, with $a, b \in [k]$. The label of node $u$ is then defined as $\ell(u) = \ell(u_a)$ where $u_a$ has the longest label $\ell$ among all nodes represented by $u$. Therefore, $\ell(u)$ is a maximal substring of $T$, and there is a bijection between the nodes in $V^T_{\fcdawg}$ and maximal substrings of $T$. Moreover, for each edge $(u, s, v) \in E^T_{\fcdawg}$, the label $s$ is a right-maximal substring of $T$, its first symbol $s[1]$ is a right-extension of the maximal string $\ell(u)$, and $\ell(u) \cdot s$ is a suffix of $\ell(v)$. In the same way as in $\stree(T)$, in $\cdawg(T)$ the edges leaving a node are lexicographically sorted by their label. In $\cdawg(T)$ there is a single node with out-degree zero, which is called the \emph{sink}, and, unlike $\stree(T)$, the leaves are the edges whose endpoint is the sink. We define $\len{u}$ as the number of leaves reachable from node $u$. We will refer to the string length of a path as the sum of the lengths of the edges in the path.

The locus $u$ of a string $P[1, m]$ in $\cdawg(T)$ is the endpoint of the path that spells out $P$ from the root. Unlike in the suffix tree, where $P$ must prefix $\ell(u)$, in a CDAWG $P$ occurs at some position $i$ in $\ell(u)$, that is, $P = \ell(u)[i, m + i - 1]$. We find $u$ by descending from the CDAWG root; the process is analogous to the descending walk from the suffix tree root. By storing the position in $T$ of the suffix $\ell(v) \cdot s$ for each leaf $(v, s, w)$ (so, $w$ is the sink), all the occurrences of $P$ can be retrieved by collecting the leaves reachable from node $u$ and adding the offset $i$.

The size $e_T$ of $\cdawg(T)$ is defined as the number of edges in $E^T_{\fcdawg}$, so $e_T$ is the number of right-extensions of maximal strings in $T$. Similarly to $E^T_{\fstree}$, the labels in the edges are not explicitly stored, but as the indices of one occurrence of the label, so $\cdawg(T)$ uses $O(e_T)$ space on top of $T$.

\begin{figure}[t]
\vspace*{-1cm}
\begin{center}
       \begin {tikzpicture}
         [-latex, auto, node distance = 2.5cm and 2.5cm, 
         on grid, semithick,
         state/.style = {circle, top color = white, bottom color = processblue!20,
        draw, processblue, text = blue, minimum width = .5 cm, 
        },
        every edge/.style={draw, draw opacity=0.3, thick},
        every node/.style={font = \scriptsize}
        ]
      \node[state] (root) {1};
      \node[state] (sink) [right = 300 pt of root] {};
      \node[state] (a) [right = of root] {2};
      \node[state] (ala) [right = of a] {3};
      \node[state] (alabar) [right = of ala] {4};
      \path (root) edge[color = brown] node {a} (a);
      \path (root) edge[dashed, bend left = 70] node[text opacity = 0.3] {\$} (sink);
      \path (a) edge[dashed, bend left = 50] node[text opacity = 0.3] {\$} (sink);
      \path (root) edge[bend left = 50, dashed] node[text opacity = 0.3] {d} (sink);
      \path (root) edge[bend left = 35] node[pos = 0.4] {bar} (alabar);
      \path (root) edge[bend left = 25] node {la} (ala);
      \path (root) edge[bend left = 50] node {r} (alabar);
      \path (a) edge node {la} (ala);
      \path (a) edge[bend left = 30pt] node {bar} (alabar);
      \path (a) edge[bend left = 25, color = brown] node[below, pos = 0.45] {r} (alabar);
      \path (ala) edge[color = brown] node {bar} (alabar);
      \path (ala) edge[bend left = 50, dashed] node[text opacity = 0.3] {l} (sink);
      \path (alabar) edge[bend left = 40, dashed] node[text opacity = 0.3] {a} (sink);
      \path (alabar) edge[dashed, color = brown] node[text opacity = 0.3] {d} (sink);
    \end{tikzpicture}

  \end{center}
    \caption{CDAWG for the string "alabaralalabarda". Dashed edges represent leaves. The root is node $1$ and the non-numbered node is the sink. Edges leading to the sink (the leaves) show only the first symbol of their label, which continues spelling the text until the \$ terminator.
    For ease to draw, edges are not shown in sorted order of labels. The edges chosen to be ``default'' to build tree $\mathcal{T}$ in Figure~\ref{fig:example-enumerator-tree} are colored in brown. }
    \label{fig:example-cdawg}
\end{figure}
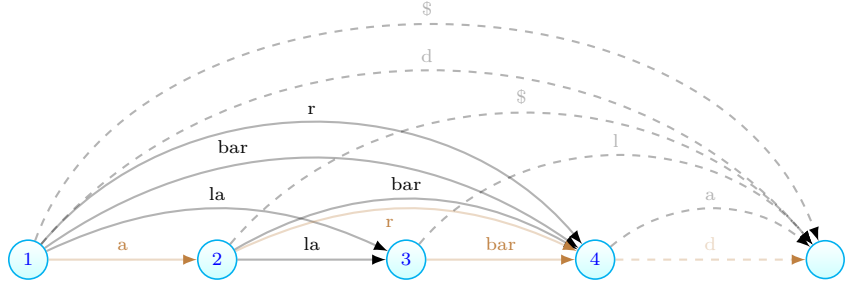

\begin{example}
Figure~\ref{fig:example-cdawg} shows the $\cdawg$ for the string "alabaralalabarda". Observe that, in the $\stree$ for the same string (Figure~\ref{fig:example-suffix-tree}), the root, node $i$ and node $iii$ have edges labeled $bar$ that reach nodes $ii$, $vi$ and $viii$, whose subtrees are isomorphic. In the $\cdawg$, there are corresponding edges labeled $bar$, reaching node $4$, from nodes $1$, $2$ and $3$. The subtrees of $iv$ and $vii$ are also isomorphic, and $\ell(iv) = r$ and $\ell(vii) = ar$ are suffixes of $bar$. The CDAWG has corresponding edges labeled $r$ to node $4$, from nodes $1$ and $2$. Thus, $\ell(4) = alabar$ and all the paths from the root to node $4$ spell out a suffix of $\ell(4)$. For example, node $4$ is the locus of $P = aba$, an $P$ (always) occurs somewhere inside $\ell(4)$. Instead, the locus of $P$ in the STree is node $vi$, and $P$ is (always) a prefix of its locus $\ell(vi)=abar$. \qed
\end{example}

\subsection{SCDAWGs}

The \emph{Symmetric}  $\cdawg$ of $T$ \cite{scdawg}, $\scdawg(T) = (V^T_{\fcdawg}, E^T_{\fscdawg})$ is the $\cdawg$ of $T$ enhanced with \textit{left-edges} $(u, s, v)$ such that $s[\len{s}]$ is a left-extension of the maximal string $\ell(u)$ and $s \cdot \ell(u)$ is a prefix of $\ell(v)$ \cite{BBHMCE87}. The left-edges leaving a node are sorted co-lexicographically by edges' labels (namely, right-to-left).

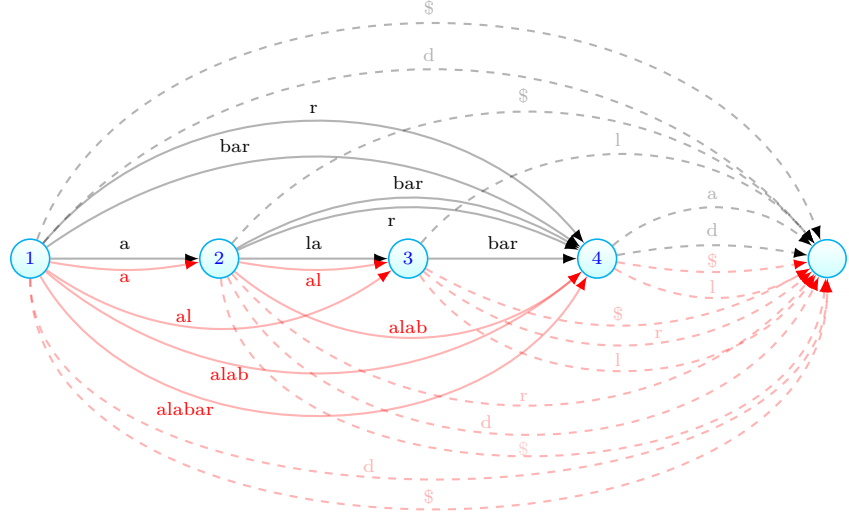
\begin{figure}[t]
\vspace*{-1cm}
\begin{center}
       \begin {tikzpicture}
         [-latex, auto, node distance = 2.5cm and 2.5cm, 
         on grid, semithick,
         state/.style = {circle, top color = white, bottom color = processblue!20,
        draw, processblue, text = blue, minimum width = .5 cm, 
        },
        every edge/.style={draw, draw opacity=0.3, thick},
        every node/.style={font = \scriptsize}
        ]
      \node[state] (root) {1};
      \node[state] (sink) [right = 300 pt of root] {};
      \node[state] (a) [right = of root] {2};
      \node[state] (ala) [right = of a] {3};
      \node[state] (alabar) [right = of ala] {4};
      \path (root) edge node {a} (a);
      \path (root) edge[dashed, bend left = 70] node[text opacity = 0.3] {\$} (sink);
      \path (a) edge[dashed, bend left = 50] node[text opacity = 0.3] {\$} (sink);
      \path (root) edge[bend left = 50, dashed] node[text opacity = 0.3] {d} (sink);
      \path (root) edge[bend left = 35] node[pos = 0.4] {bar} (alabar);
      \path (root) edge[bend left = 50] node {r} (alabar);
      \path (a) edge node {la} (ala);
      \path (a) edge[bend left = 30pt] node {bar} (alabar);
      \path (a) edge[bend left = 25] node[below, pos = 0.45] {r} (alabar);
      \path (ala) edge node {bar} (alabar);
      \path (ala) edge[bend left = 50, dashed] node[text opacity = 0.3] {l} (sink);
      \path (alabar) edge[bend left = 40, dashed] node[text opacity = 0.3] {a} (sink);
      \path (alabar) edge[bend left = 10, dashed] node[text opacity = 0.3] {d} (sink);
      
      \path (root) edge[color = red, bend right = 10] node[yshift = 2, below] {a} (a);
      \path (root) edge[bend right = 40, color = red] node[below, pos = 0.35, yshift = 2] {alab} (alabar);
      \path (root) edge[color = red, bend right = 35] node[above, pos = 0.4, yshift = -2] {al} (ala);
      \path (root) edge[bend right = 60, color = red] node[below, pos = 0.3] {alabar} (alabar);
      \path (root) edge[dashed, in = 270, out = 270, distance = 115, color = red, yshift = -2] node[yshift = -2, above, text opacity = 0.3] {\$} (sink);
      \path (root) edge[in = 270, out = 270, distance = 100, dashed, color = red] node[pos = 0.45, text opacity = 0.3, above, yshift = -2] {d} (sink);

      \path (a) edge[dashed, bend right = 85, color = red] node[above, yshift = -2, text opacity = 0.2, yshift = -2] {\$} (sink);
      \path (a) edge[dashed, bend right = 65, color = red] node[above, yshift = -2, text opacity = 0.3, pos = .45] {d} (sink);
      \path (a) edge[dashed, bend right = 50, color = red] node[above, text opacity = 0.3, pos = 0.5, yshift = -2] {r} (sink);
      \path (a) edge[color = red, bend right = 40] node[above, pos = .5, yshift = -2] {alab} (alabar);
      \path (a) edge[color = red, bend right = 10] node[below, yshift = 2] {al} (ala);

      \path (ala) edge[dashed, bend right = 30, color = red] node[above, text opacity = 0.3, yshift = -2] {\$} (sink);
      \path (ala) edge[dashed, bend right = 40, color = red] node[above, text opacity = 0.3, pos = .6, yshift = -2] {r} (sink);
      \path (ala) edge[color = red, bend right = 55, dashed] node[above, text opacity = 0.3, yshift = -2] {l} (sink);

      \path (alabar) edge[dashed, bend right = 10, color = red] node[above, text opacity = 0.3, yshift = -2] {\$} (sink);
      \path (alabar) edge[color = red, bend right = 30, dashed] node[above, text opacity = 0.3, yshift = -2] {l} (sink);

    \end{tikzpicture}

    \vspace*{-1cm}
  \end{center}
    \caption{SCDAWG for the string "alabaralalabarda". Black edges represent right-edges, and red edges represent left-edges. Dashed edges represent leaves. The root is node $1$ and the non-numbered node is the sink. Right-edges leading to the sink show only the first symbol of their label, which continues spelling the text until the \$ terminator, and left-edges leading to the sink show only the last symbol of their label, which continues spelling the text to the left until the first symbol of the string (assuming a \$ terminator to the left of it).
    For ease to draw, edges are not shown in sorted order of labels.}
    \label{fig:example-scdawg}
\end{figure}

For a given left-edge $(u, s, v)$ $\in E^T_{\fscdawg}$, since $\ell(u)$, $\ell(v)$ are maximal strings, then in $\cdawg(\rev{T})$ there exist exactly two nodes $u_l$, $v_l$ such that $\ell(u_l) = \rev{\ell(u)}$, $\ell(v_l) = \rev{\ell(v)}$ and $\ell(u_l) \cdot \rev{s}$ is a suffix of $\ell(v_l)$, so in $E^{\rev{T}}_{\fcdawg}$ there is an edge $(u_l, \rev{s}, v_l)$. Then, there is a one-to-one correspondence between nodes in $\cdawg(T)$ and $\cdawg(\rev{T})$, and the left-edges in $E^T_{\fscdawg}$ are exactly the edges in $E^{\rev{T}}_{\cdawg}$ (yet, note that the edge labels are $s$, not $\rev{s}$). It naturally follows that the size of $\scdawg$ is $\overline{e_T} = e_T + e_{\rev{T}}$. Figure~\ref{fig:example-scdawg} shows the $\scdawg$ for the string "alabaralalabarda".

$\scdawg(T)$ can be built in $O(n \log \sigma)$ time by computing $\cdawg(T)$ and $\cdawg(\rev{T})$ in $O(n)$ time \cite{scdawg} and mapping each node by following the path of each maximal string $s$ and $\rev{s}$ in $\cdawg(T)$ and $\cdawg(\rev{T})$, respectively. Further, Blumer et al.~\cite{BBHMCE87} give an $O(n)$-time construction, and Inenaga et al.~\cite{scdawg} give an online linear-time algorithm.

\section{Optimal Contextual Pattern Matching in $O(\overline{e})$ Space}

\subsection{Sub-optimal solution} \label{sec:subopt}

Let $u$ be the CDAWG node that is the locus of $P$ and let $h \ge m$ be, initially, the string length of the path we followed from the CDAWG root to $u$. Note that $u$ itself is the only right-context locus if $h \ge m+\lambda$, so below we assume the non-trivial case, where we still need to match $m+\lambda-h > 0$ symbols from the right context. The loci of the right-contexts of $P$ are then all the closest descendants $v$ of $u$ reached by paths of string length $\ge m+\lambda-h$. So, for each right-edge $(u, s, v)$ we must evaluate if $m + \lambda \leq h + \len{s}$, in which case $v$ is the locus of a right-context of $P$, and $P$ occurs at position $|\ell(v)| - (h + |s|) + 1$ of $\ell(v)$. Otherwise, we continue recursively looking from right-context loci from $v$, updating $h$ to $h+|s|$.

Let us now find the left-contexts for each right-context found at a node $v$. Note that we have reached $v$ by a path of length $h \ge m+\lambda$, which matches $P$ and a right-context, so $h$ is the length of the suffix of $\ell(v)$ starting with the occurrence of $P$. If $\len{\ell(v)} - h \ge \lambda$, then, all those right-contexts occur with the same left-context and we are done. Otherwise, we set $h' = |\ell(v)|-h$ and start traversing left-edges to find left-contexts. For each left-edge $(v, t, w)$ we evaluate if $\lambda \leq h' + \len{t}$ to determine if we have matched a left-context of $P$.

Note that, if for some right-edge (resp., left-edge), it happens that the destination is the sink, then this edge is a leaf and this contextual occurrence does not share a length-$\lambda$ right-context (resp., left-context) with any other occurrence of $P$, so we can report it immediately. 

\begin{example}\label{example:right-left-contexts}
In Figure~\ref{fig:example-scdawg}, for the right-edge $(2, bar, 4)$ we have $\ell(2) = a$, $\ell(4) = alabar$, and $\ell(2) \cdot bar$ is a suffix of $\ell(4)$. Note also that following this right-edge extends the contexts of the string $a$ to the right and to the left, i.e., we get $al \cdot a \cdot bar$.

Similarly, following the left-edge $(2, alab, 4)$ we have that $alab \cdot \ell(2)$ is a prefix of $\ell(4)$. Further, following this edge extends the contexts of the string $a$ to the right and to the left, i.e., we get $alab \cdot a \cdot r$. \qed
\end{example}

Algorithm~\ref{alg:contextual} gives the procedure to report all the contextual occurrences of $P$ once we have found its locus $u$ and the string length $h$ of the path followed to $u$. It uses $O(\overline{e}_T)$ space by using a {\scdawg} augmented with some fields in nodes and edges:
For every node $u \in V^T_{\fscdawg}$, we store the length of $\ell(u)$ and the position $u.occ$ of one occurrence of $\ell(u)$ in $T$. Additionally, for every right-edge $e = (u, s, v)$, we store the edge length $\vert s \vert$, and if $e$ is a leaf, we store the position $e.occ$ of the suffix of $T$ that $e$ represents. We store analogous fields for left-edges. 

\begin{algorithm}[t]\label{alg:contextual}
\SetKwFunction{fnContextual}{contextual}
\SetKwFunction{fnLeftContextual}{leftContextual}
\Function{\fnContextual{$u,h$}}{
  \lIf{$h \ge m+\lambda$}{
     $\fnLeftContextual(u,\len{\ell(u)}-h)$
     }
  \Else {
        \For{{\bf each} {\rm right-edge} $e = (u,s,v)$}{
            \lIf{$v$ is the sink}{\textbf{report} $e.occ + (\len{\ell(u)} - h + 1)$};
            \textbf{else} $\fnContextual(v,h+|s|)$\;}
        }
}

\bigskip
\Function{\fnLeftContextual{$v,h'$}}{
  \lIf{$h' \ge \lambda$}{
     \textbf{report} $v.occ+(h'+1)$}
  \Else {
        \For{{\bf each} {\rm left-edge} $e = (v,t,w)$}{
            \lIf{$w$ is the sink}{\textbf{report} $e.occ + (h' + 1 + \len{t})$}
            \textbf{else} $\fnLeftContextual(w,h'+|t|)$\;}
        }
}
\caption{Function \texttt{contextual($u,h$)} reports all the  $\lambda$-contextual occurrences of $P[1,m]$ from its locus $u$ and the string length $h$ of the path followed to $u$. Function \texttt{leftContextual($v,h'$)} reports all the $\lambda$-contextual occurrences reachable by traversing $\lambda-h'$ symbols labeling left-edges from $v$. The fields $e.occ$ and $v.occ$ are text positions where the strings represented by leaves $e$ or nodes $v$, respectively, occur in $T$. We report the positions where $P$ occurs.}
\end{algorithm}

To find the locus of $P[1,m]$, we descend by following right-edges while matching substrings of $P$ in $\scdawg$. We start by setting $u$ to the $\scdawg$ root, use a variable $h$ (initially set to $0$) to mark the part of $P$ already compared.
We binary search right-edges of $u$ to find the edge $e = (u, s, v)$ such that $s[1] = P[h+1]$. If neither $s$ is a prefix of $P[h+1, m]$ or vice-versa, then $P$ does not occur in $T$. Otherwise, we set $h \gets h + \len{s}$ and $u \gets v$. We repeat this process until $m \leq h$, so $u$ is the locus of $P$. Then we call \texttt{contextual($u,h$)} of Algorithm~\ref{alg:contextual}.

The search for the locus of $P$ takes $O(m \log \sigma)$ time. After finding it, for each edge traversed from a node to all its children we increment the number of contextual occurrences to report, so it takes $O(occ)$ time to find all $occ$ contextual occurrences of $P$.\footnote{As an alternative amortized argument, note that the tree of recursive calls to \texttt{contextual} and \texttt{leftContextual} reports one occurrence at each leaf and has at least two children per node, so there are less than two calls per occurrence reported.} The overall time is thus $O(m \log \sigma + occ)$. The space usage is $O(\overline{e}_T)$ on top of $T$, but it still needs $T$ to compare $P$ with the edges of $\scdawg$, so we use $O(\overline{e}_T + n) = O(n)$ space. We next solve this space issue, and also improve the locus finding time.

\subsection{Optimal-time solution} \label{sec:opt}

We now improve the above solution in both time and space. Belazzougui and Cunial~\cite{belazzouggui17} give a data structure of size $O(e_T)$ that determines if a pattern $P[1, m]$ occurs in a string $T[1, n]$ in $O(m)$ time. Using this structure, we first determine whether $P$ occurs in $T$. If $P$ does not occur in $T$, then there are no contextual occurrences. Otherwise, we search for the locus of $P$ through a \textit{blind search} \cite{belazzouggui17}. To do this, we add at each edge $e = (u, s, v)$ of our data structure the first character of the label $s$ in a variable $e.label = s[1]$ (we now do not store $s$, but still store $|s|$). Now, with $u$ starting from the root and $h=0$, we follow the edge $e$ of $u$ such that $e.label = P[h+1]$ (we use perfect hashing \cite{FKS84} to determine which edge to follow in $O(1)$ time), and update $h$, and $u$ as before. Since we know that $P$ occurs in $T$, we can skip comparing the whole label of $e$ against $P[h+1, m]$. We continue that way until $m \leq h$, which means $u$ is the locus of $P$. With this change, the overall time is now $O(m + occ)$ and we can drop the text $T$ (because labels are not needed anymore).

\begin{theorem} \label{thm:cpm}
The $occ$ solutions to Contextual Pattern Matching for $P[1,m]$ can be listed in optimal time $O(m+occ)$, on a data structure of size $O(\overline{e})$.
\end{theorem}

\section{Enumerators for Contextual Pattern Matching}

A standard traversal of the $occ$ results, though taking $O(occ)$ time, may spend $O(\lambda)$ time between some result and the next.
We now describe an enumerator that, after $O(m)$ time preprocessing, can deliver each new solution to the Contextual Pattern Matching problem within $O(\log\log \lambda)$ time. 

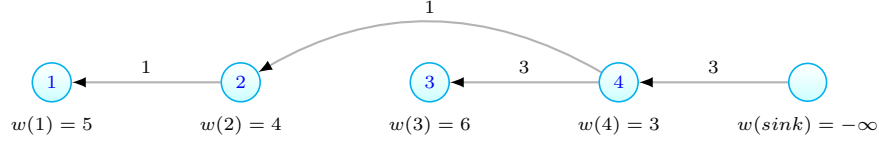
\begin{figure}[t]
\begin{center}
       \begin {tikzpicture}
         [-latex, auto, node distance = 2.5cm and 2.5cm, 
         on grid, semithick,
         state/.style = {circle, top color = white, bottom color = processblue!20,
        draw, processblue, text = blue, minimum width = .5 cm, 
        },
        every edge/.style={draw, draw opacity=0.3, thick},
        every node/.style={font = \scriptsize}
        ]

      \node[state] (alabar) [right = of ala] {4};
      \node[state] (sink) [right = of alabar] {};
      \node[state] (a) [right = of root] {2};
      \node[state] (ala) [right = of a] {3};
      \node[state] (root) {1};

      \node () [below = of alabar, yshift = 55pt] {$w(4) = 3$};
      \node () [below = of ala, yshift = 55pt] {$w(3) = 6$};
      \node () [below = of a, yshift = 55pt] {$w(2) = 4$};
      \node () [below = of root, yshift = 55pt] {$w(1) = 5$};
      \node () [below = of sink, yshift = 55pt] {$w(sink) = -\infty$};

      \path (a) edge node[above] {1} (root);
      \path (alabar) edge[bend right] node[above] {1} (a);
      \path (alabar) edge node[above] {3} (ala);
      \path (sink) edge node[above] {3} (alabar);
    \end{tikzpicture}

  \end{center}
    \caption{A possible enumerator tree $\mathcal{T}$ for the right-edges of the $\scdawg$ of the string "alabaralalabarda". The default edges chosen are those colored in brown in Figure~\ref{fig:example-cdawg}. Edges $(u, s, v)$ are labeled by their edge length $\len{s}$.
    }
    \label{fig:example-enumerator-tree}
\end{figure}

The idea is inspired by a recent result by Adamson et al.~\cite{AGM24}, for enumerating the paths of a certain length in a graph. Let us first describe our solution on CDAWGs for simplicity. We arbitrarily choose exactly one (right-)edge leaving each (non-sink) node as its {\em default} edge. The set of all default edges then defines an ``enumerator'' tree $\cal T$, where the target of the default edge is the parent of the source and the CDAWG sink is the root of $\cal T$. The nodes $u$ of $\cal T$ have weight $w(u)$ defined as the sum of edge lengths from the root of $\cal T$ to $u$. The size of $\cal T$ is the number of CDAWG nodes. Figure~\ref{fig:example-enumerator-tree} shows one possible enumerator tree for the right-edges of the $\scdawg$ of the string "alabaralalabarda".

Let $u$ be the CDAWG node that is the locus of $P$, found in $O(m)$ time as explained in Section~\ref{sec:opt}. Let $h \ge m$ be the string length of the path we followed from the CDAWG root to $u$, computed as in Section~\ref{sec:subopt}. Our first task is, again, to enumerate the loci of the right-contexts of $P$, that is, the closest descendants $v$ of $u$ by paths of string length $\ge m+\lambda-h$.

We will first find one of those loci, $u^*$, which is reached from $u$ by default edges, that is, $u^*$ is an ancestor of $u$ in $\cal T$. Node $u^*$ is reached from $u$ in $\cal T$ using a {\em weighted ancestor query} on the node weights: $u^*$ is the lowest ancestor of $u$ such that $w(u)-w(u^*) \ge m+\lambda-h$, or equivalently, $w(u^*) \le w(u) - (m+\lambda-h)$.\footnote{To capture contexts shorter than $\lambda$ near the end of $T$, we set $w(\cdot) = -\infty$ for the sink node.} We show in Section~\ref{sec:weighted} how to solve this query in time $O(\log\log\lambda)$, by exploiting the fact that the difference between $w(u)$ and the weight of the child of $u^*$ is at most $\lambda$.

\begin{algorithm}[t!] 
\SetKwFunction{fnRetrieve}{enumerate}
\SetKwFunction{fnLeftRetrieve}{leftEnumerate}
\SetKwFunction{fnWancestor}{wancestor}
\Function{\fnRetrieve{$u,h$}}{
  \lIf{$h \ge m+\lambda$}{
     $\fnLeftRetrieve(u,|\ell(u)|-h)$
     }
  \Else {
     $u^* \gets \fnWancestor(\mathcal{T},u,w(u)-(m+\lambda-h))$ ; \\
     $\fnLeftRetrieve(u^*,|\ell(u^*)|-(h+w(u^*)-w(u)))$ ;\\
     $x \gets u$ ;\\
     \While{$x \neq u^*$}{
        $(x,s',y') \gets$ default right-edge leaving $x$ ;\\
        \For{{\bf each} {\rm right-edge} $(x,s,y) \neq (x,s',y')$}{
            $\fnRetrieve(y,h+|s|)$ ;}
        $x \gets y'$ ; \\
        $h \gets h+|s'|$ ;
      }
   }
}
\bigskip
\Function{\fnLeftRetrieve{$v,h'$}}{
  \lIf{$h' \ge \lambda$}{
     \textbf{report} $v.occ + (h'+1)$
     }
  \Else {
     $v^* \gets \fnWancestor(\mathcal{T}',v,w(v)-(\lambda-h'))$ ; \\
     \textbf{report} $v^*.occ +(h'+w(v^*)-w(v)+1)$;\\
     $y \gets v$ ;\\
     \While{$y \neq v^*$}{
        $(y,t',z') \gets$ default left-edge leaving $y$ ;\\
        \For{{\bf each} {\rm left-edge} $(y,t,z) \neq (y,t',z')$}{
            $\fnLeftRetrieve(z,h'+|t|)$ ;}
        $y \gets z'$ ; \\
        $h' \gets h'+|t'|$ ;
      }
   }
}

\caption{Iterator \texttt{enumerate($u,h)$} enumerates all the $\lambda$-contexts from the locus $u$ of $P[1,m]$; $h \ge m$ is the string length of the path to the locus. Auxiliary enumerator \texttt{leftEnumerate($v,h'$)} enumerates all the left-contexts reachable by traversing $\lambda-h'$ symbols from $v$. $\mathcal{T}$ and $\mathcal{T}'$ are the enumerator trees for right- and left-edges, respectively, and $w$ is the node weight in those trees. Keyword {\bf report} enumerates each new context; we report the positions where $P$ occurs.}
  \label{alg:enumerate}
\end{algorithm}

Once we have obtained our first locus, $u^*$, we {\em visit} all of its CDAWG ancestors $x$ by default edges, from $x=u$ to (but not including) $u^*$. {\em Visiting} $x$ means descending from $x$ by all the non-default edges $(x,s,y)$ and repeating the process recursively from every such $y$, by regarding $y$ in $\cal T$, performing a weighted ancestor query to find a locus $y^*$, and {\em visiting} all the nodes in the path from $y$ to the parent of $y^*$.

Along this recursive procedure, we deliver each new locus node $v$ with delay $O(\log\log\lambda)$: our recursion first performs an $O(\log\log\lambda)$-time weighted ancestor query and immediately enumerates the locus $u^*$. Then it visits all the nodes $x$ in the path between $u$ and $u^*$. There is exactly one default child of each node $x$ and at least one non-default child, so all the non-default children $y$ of all nodes $x$ in the path are found with constant delay. Further, there is at least one other occurrence to report below each such $y$. The argument then applies inductively in the subtree rooted at each child $y$ (we find a first occurrence $y^*$ in $O(\log\log\lambda)$ time by the default path, report it, visit the nodes in between, and so on).

\begin{example}
Consider, in Figure~\ref{fig:example-cdawg}, the process of finding the right-contexts of $P=$"a" with $\lambda=2$. We arrive at node $u=2$ with $h=1$. A weighted ancestor query (see Figure~\ref{fig:example-enumerator-tree}) for the closest node at distance $\ge m+\lambda-h=2$ yields $u^*$ as the sink, where we report the right-context "rd". We now traverse the path in $\cal T$ from $u$ to $u^*$, visiting the nodes $2$ and $4$. Node $2$ has three other right-edges that do not lead to node $4$ by "r" (the default edge): one leads to the sink by "\$", another to node $3$ by "la", and the third to node $4$ by "bar". The first case yields the empty right-context corresponding to the final "a" in the text. In the other cases we exceed the context length $\lambda$ immediately, yielding the contexts "la" and "ba", respectively. We now descend by the default edge to node $4$ by "r". From it, we descend by the only non-default edge to the sink, by "a", leading to the remaining right-context, "ra".   \qed
\end{example}

To complete the process, we must obtain all the left-contexts that correspond to each right-context. Those are obtained with essentially the same technique, now using the left-edges of the SCDAWG. We build another enumerator tree, ${\cal T}'$, on the left-edges, and at query time repeat the process from the locus $v$ of each new right-context. If the path followed to $v$ was of string length $h \ge m+\lambda$, then the left-context loci are the closest nodes reached from $v$ by left-edge paths of string length $\ge \lambda-(\ell(v)-h)$. When all the left-contexts of $v$ are reported, we switch to the next right-context. The total delay is still $O(\log\log\lambda)$.

Algorithm~\ref{alg:enumerate} gives the pseudocode. Though we describe recursive procedures for enumeration, those are easily turned into iterators by maintaining the state of the computation and returning to the caller with each new result. The memory required by the enumeration is $O(\lambda)$, as $\lambda$ is the maximum height of the recursion stack (just as with the standard traversal procedure).

\begin{theorem} \label{thm:enumerate}
The solutions to Contextual Pattern Matching for $P[1,m]$, with context length $\lambda$, can be enumerated with $O(m)$ preprocessing time and $O(\log\log\lambda)$ delay, on a data structure of size $O(\overline{e})$. The enumerator requires $O(\lambda)$ space.   
\end{theorem}

\section{Distance-Sensitive Weighted Ancestor Queries }
\label{sec:weighted}

The weighted ancestor problem on trees assumes that nodes have integer weights $w(\cdot)$ that are increasing towards descendants. Given a tree node $u$ and a target weight $t$, the query finds the lowest ancestor $u^*$ of $u$ such that $w(u^*) \le t$. It has been shown that this problem can be reduced essentially to (a constant number of) predecessor queries \cite{FM96,KL07}, and thus it can be solved in $O(n)$ space and $O(\log\log n)$ time on a tree of $n$ nodes.

We are interested in exploiting the fact that $w(u)-t$ may be small (it is at most $\lambda$ in the preceding section), aiming at time $O(\log\log(w(u)-t))$. This does not immediately stem from previous work \cite{KL07}, as they include an additive term of the form $O(\log^* n)$, necessary to achieve linear space. We follow their basic scheme, using a distance-sensitive predecessor data structure, and use a different way to achieve linear space (Gawrychowski et al.~\cite{gawrychowski2014weighted} give an alternative way to remove this $O(\log^* n)$ additive term using micro/macro trees, yet still do not aim at the distance-sensitive time we obtain in this section).

We partition the tree into paths; those can be heavy paths \cite{FM96} or centroid paths \cite{KL07} indistinctly for us. The important property is that every node reaches the root with only $O(\log n)$ changes from one path to another. Each node stores the weights of the highest nodes in each of those paths that connect it to the root. A first part of the search for $u^*$ is to determine in which of those paths it lies. Since this is a predecessor query for $t$ on $O(\log n)$ integer weights, the correct path $\pi$ is found in constant time using fusion trees \cite{FW93}.

On each path $\pi$ we build a linear-space distance-sensitive predecessor data structure \cite{BBV12}, so we again look for the predecessor of $t$ and finish. The query time of this data structure is $O(\log\log\Delta)$, where $\Delta$ is the minimum between $t-w(u^*)$ and $w(u')-t$, where $u'$ is the child of $u^*$ in its path to $u$. Since $w(u')-t \le w(u)-t$, we have the desired time complexity.

The problem of this scheme is that its space complexity is $O(n\log n)$. Just as in previous work \cite{KL07}, we prune the tree, turning into leaves the lowest nodes whose subtree has more than $\log n$ nodes. The pruned tree then has $O(n/\log n)$ nodes and we can afford to use the general scheme on it within $O(n)$ space. On the small detached subtrees, with root $x$, we first check if $w(x) > t$, in which case we rerun the query from the leaf $x$ that corresponds to in the main tree. If $w(x) \le t$, instead, we must run the query on the small subtree. 

To do this efficiently, we repeat the same scheme on the small trees. Now there are $O(\log(\log n))$ paths between a node and its root, so the extra space is $O(n\log\log n)$. To maintain linear space, we prune those small subtrees, turning into leaves their nodes whose subtree sizes exceed $\log\log n$. We now must show how to solve queries on the tiny detached subtrees of size $O(\log\log n)$.

Instead of continuing this process for $O(\log^* n)$ rounds \cite{KL07}, we resort to universal tables at this point. Let $k = O(\log\log n)$ be the maximum size of the tiny trees. We remap their weights to the interval $[1,k]$, and store a fusion tree on those $k$ elements that allows us map $t$ to the corresponding value in constant time. Now, $2k$ bits suffice to describe the tree topology \cite{MR01} and $k\lceil \log_2 k\rceil$ bits suffice for the weights. There are $k^2$ possible weighted ancestor queries ($k$ nodes and $k$ target weights). Therefore, a table of $2^{k(2+\lceil \log_2 k\rceil)} \times k^2 = o(n)$ entries can store all the precomputed answers to queries on every possible tiny tree.

\begin{theorem} \label{thm:weighted}
Weighted ancestor queries $(u,t)$ on a tree of $n$ nodes with integer weights $w(\cdot)$ can be solved in time $O(\log\log (w(u)-t))$ using an $O(n)$-space data structure. 
\end{theorem}

\section{Conclusions and Future Work}

We have presented the first compressed index (for highly repetitive text collections) that supports contextual pattern matching in optimal time, $O(m+occ)$. The space, $O(\overline{e})$, is proportional to the size of the SCDAWG of the text. We have also presented an enumerator for the results, which after $O(m)$ time delivers each new contextual occurrence in time $O(\log\log\lambda)$, for a context length $\lambda$. In our way, we uncover a new, improved, linear-space distance-sensitive data structure for weighted ancestor queries on trees.

One interesting challenge is to obtain constant delay, which can be achieved by running weighted ancestor queries in constant time. This has been shown to be possible on suffix trees \cite{gawrychowski2014weighted,belazzougui2021weighted}; the results  could perhaps be translated to our enumerator trees $\cal T$ and ${\cal T}'$. Another interesting challenge is to solve contextual pattern matching, even if not in optimal time, within space bounded by a stronger measure of repetitiveness, such as the size of a context-free grammar representing the text.

\section*{Acknowledgements}

We thank the reviewers for their careful reading and detailed comments to improve our presentation.

%
%
%
\bibliographystyle{splncs04}
\bibliography{biblio}

\end{document}